\documentclass[aps,prl,reprint,amsmath,amssymb,superscriptaddress,longbibliography]{revtex4-1}

\usepackage{graphicx}
\usepackage{dcolumn}
\usepackage{bm}
\usepackage{amsmath}
\usepackage{color}
\usepackage{xspace}
\usepackage{ulem}
\usepackage{xfrac}
\usepackage{color,soul}
\usepackage{graphicx}
\usepackage[colorlinks=true,urlcolor=blue,citecolor=blue,linkcolor=blue,bookmarks=false,pdfstartview={FitH}]{hyperref}

\newcommand{\dg}{^{\dagger }}

\newcommand{\bk}{{\bf k}}
\newcommand{\bq}{{\bf q}}

\begin{document}

\title{Superconductivity from energy fluctuations in dilute quantum critical polar metals}

\author{Pavel A. Volkov}
\email{pv184@physics.rutgers.edu}
\affiliation{Center for Materials Theory, Department of Physics and Astronomy, Rutgers University, Piscataway, NJ 08854, USA}
\author{Premala Chandra}
\affiliation{Center for Materials Theory, Department of Physics and Astronomy, Rutgers University, Piscataway, NJ 08854, USA}
\author{Piers Coleman}
\affiliation{Center for Materials Theory, Department of Physics and Astronomy, Rutgers University, Piscataway, NJ 08854, USA}
\affiliation{Department of Physics, Royal Holloway, University of London, Egham, Surrey TW20 0EX, UK}
\date{\today}

\begin{abstract}
Superconductivity in low carrier density metals challenges the
conventional electron-phonon theory due to the absence of retardation
required to overcome Coulomb repulsion. 
In quantum critical polar metals, the Coulomb repulsion is heavily screened, while the critical transverse optic phonons decouple from the electron charge. In the resulting vacuum, the residual interactions between quasiparticles are carried by energy fluctuations of the polar medium, resembling the gravitational interactions of a dark matter universe. Here we demonstrate that
pairing inevitably emerges from "gravitational'' interactions with the energy fluctuations,
leading to a dome-like 
dependence of the superconducting $T_c$ on carrier density. Our estimates show that this mechanism may explain the critical temperatures observed in doped SrTiO$_3$. We provide predictions for the enhancement of superconductivity near polar quantum criticality in two and three dimensional materials that can be used to test our theory.
\end{abstract}

\maketitle

Superconductivity exemplifies the dramatic effects
of interactions in many-body quantum systems
\cite{cooper1956}. 
Conventional
superconductors 
electrons exploit  
the electron-phonon attraction to overcome the Coulomb repulsion 
\cite{tolmachev1958,morel1962} by producing  a highly retarded 
attraction that pairs electrons, 
a process that requires a large ratio
between the  Fermi and Debye energies 
  $E_{F}/\omega_{D}>>1$ \cite{gurevich1962}.
A challenge to this mechanism is posed by superconductivity in low
carrier polar metals. These lightly doped insulators, exemplified
by doped SrTiO$_3$ (STO) \cite{collignon2019} lie close to a
ferroelectric quantum critical point (QCP) and exhibit bulk superconductivity down to  carrier densities of order $10^{19}cm^{-3}$, where the relevant phonon frequency exceeds the Fermi energy \cite{collignon2019}
by orders of magnitude.
Yet despite this inversion of energy scales, experiments \cite{lin2015,swartz2018}
indicate a conventional s-wave condensate, with a ratio of gap to
transition temperature $2\Delta /T_{c}\approx 3.5$ in agreement
with BCS theory \cite{swartz2018}. 

Several theories have been advanced to explain superconductivity in
polar metals using conventional electron-phonon interaction \cite{gorkovpnas} and its extension to include plasmon effects
\cite{takada1978,takada1980,ruhman2016,gorkov2017,enderlein2020,Ma2021}. 
Alternative phonon coupling mechanisms requiring spin-orbit coupling
or multiband effects \cite{yanase2018,gastiasoro2020,kanasugi2020,volkov2020} have
also been examined. 
Recently, it has been proposed that the 
underlying ferroelectric quantum criticality of the polar
metal is a key driver in the pairing
\cite{edge.2015,stucky2016,rischau.2017,tomioka2019}. 
However, this appealing idea encounters a difficulty, for the critical modes
of a polar QCP are transverse optic (TO) phonons, 
which are neutral and decouple from the electrons\cite{wolfle.2018,wolfle.2018com,wolfle.2018repl}. 

These considerations motivate us to reconsider
\begin{figure}[h]
	\includegraphics[width=0.5\textwidth]{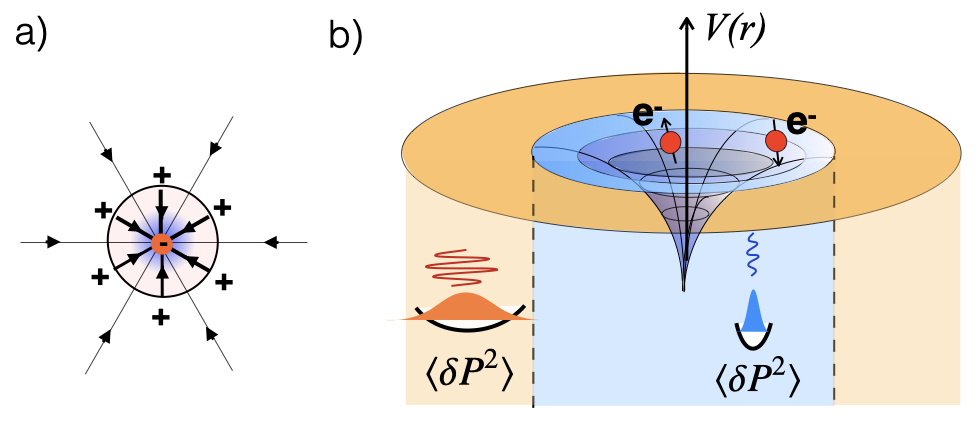}
	\caption{Interactions between electrons in a quantum critical
	polar metal: (a) the electric lines of force 
around an electron are ionically  screened,
(b) the fluctuations of the phonon energy density around electrons (see \eqref{eq:h2ph}) create an attractive potential well.
}
	\label{fig:cartoon}
\end{figure}
superconductivity in quantum critical polar metals, guided by
two key observations: first, the strong ionic screening associated with the enhanced 
dielectric constant severely weakens the electronic Coulomb
interaction (Fig \ref{fig:cartoon} (a)); second, since the critical transverse optic
phonon modes decouple from the electron charge, the
resulting quantum fluid can be likened to 
dark matter, for
like baryons in the cosmos, electron charge
does not directly interact with the 
the intense background of zero-point
fluctuations. Moreover, like dark matter, 
the quantum critical TO modes are only revealed to the electrons 
via their ``gravitational interaction'', mediated by the stress-energy tensor. The resulting interaction becomes increasingly
intense at a quantum critical point;
we model it by the Hamiltonian 
\cite{ngai1974,maslov2020}:
\begin{equation}
H_{\rm En} = g
\int d^{3}x \rho_{e} ({\bf x})\vec P ({\bf x})^{2}
\label{eq:h2ph}
\end{equation}
where $\rho_{e} ({\bf x})= \psi \dg ({\bf x})\psi (x)$ is the electron density, $(\vec P ({\bf x}))^{2}$ is proportional to the energy density
of the local polarization $\vec{P}$ and $g$ is a coupling constant
with the dimensions of volume. This coupling locally suppresses 
the zero-point fluctuations of the polarization
in the vicinity of electrons, that
in turn lowers the chemical potential of
electrons passing in the region (Fig. \ref{fig:cartoon} (b)), creating an attractive
potential well. 
To lowest order, the resulting attractive potential is described by the virtual exchange of pairs of TO phonons \footnote{See Supplemental Material at [URL will be inserted by publisher] for the details.},
allowing us to link
these ideas to two recent observations:
first,  that two-phonon exchange appears to drive the anomalous
``high-temperature'' $T^{2}$ resistivity of polar metals\cite{maslov2020,nazaryan2021}, and
second that two-phonon processes may drive superconductivity
\cite{marel2019} reviving an old idea \cite{ngai1974}.

Here we study the consequences of the coupling to energy fluctuations
of the order parameter \eqref{eq:h2ph} in quantum critical polar
metals. We find that quantum criticality causes 
the zero-point fluctuations to intensify, increasing the attractive
interaction between electrons. 
Moreover, while the  Fermi liquid behavior is robust 
against these couplings, the attractive
interactions mediated by the energy fluctuations 
overcome the Coulomb repulsion in the
low density regime. Using model parameters for SrTiO$_3$ we find agreement with experiments both in the magnitude of $T_c$ and in its doping 
dependence; finally we predict the
fingerprints of this novel mechanism to be more pronounced in
two-dimensional quantum critical polar systems.

Our theory is built on an isotropic model for the polar metal, with an
action $S= S_{e}+S_{C}+S_{\rm En}$, where $S_{\rm En}= \int d\tau  H_{\rm
En}$ is the energy fluctuation term (\ref{eq:h2ph}), $S_{e}=\sum_k\psi\dg_{k}(\epsilon_{\bk}-i\omega_n)\psi_{k}$,
is the electronic action, in terms of the Fourier transformed electron
field $\psi_{k}$, where 
$k\equiv (i\omega_{n},\bk
)$ is a four-vector containing the Matsubara frequency and wavevector
$\vec{k}$ and 
\begin{equation}\label{coulomb1}
S_{C}= \sum_{q} \left[\frac{|e\rho_{q}- (\nabla \cdot
\vec{P})_{q}|^{2}}{2\varepsilon_{0}\varepsilon_{1}\vec{q}^{2}} + 
\frac{\nu_n^2+\omega_{T}^{2}(\vec{q})}
{2\varepsilon_0 \Omega_{0}^{2}}
|\vec{P} (q) |^{2} 
\right],
\end{equation}
is the Coulomb interaction between 
the total charge densities
$e\rho_{e}-\nabla \cdot \vec{P}$, where
$q\equiv (i\nu_n,\vec{q})$.
$\varepsilon_1$ is the bare dielectric constant, $\Omega_{0}$ is the ionic plasma
frequency and $\omega_{T}(\vec{q})$ is the energy dispersion of the transverse optical mode. At low momenta $\omega_{T}^2(\vec{q}) \approx \omega_{T}^{2}+ c_s^{2}\vec q^{2}$, where $c_s$ is the speed of the transverse optic mode and $\omega_{T}^{2}$ vanishes at the QCP. The Gaussian coefficients of the polarization,  $\delta^{2}S_{C}/\delta
P_{a} (-q)\delta P_{b} (q)=D^{-1}_{ab} (q) $
in \eqref{coulomb1},
separate into 
transverse and longitudinal components
\begin{equation}\label{}
D^{-1}_{\alpha \beta }(q) 
= D^{-1}_{L} (q)\hat q_{\alpha }\hat q_{\beta }
+ D^{-1}_{T} (q) (1-\hat q_{\alpha }\hat q_{\beta })
\end{equation}
where $D^{-1}_{L,T} (q)=  (\nu_n^2+\omega_{T,L}^{2}(\vec{q}))/\varepsilon_{0}\Omega_{0}^{2}
$  are the inverse longitudinal and transverse phonon propagators. 
The longitudinal optic mode frequency $\omega_{L}^{2}(\vec{q})= \omega_{T}^{2}+ \Omega_{0}^{2}/\epsilon_{1}$ is shifted upwards by the Coulomb interaction.

We first consider the case where $g=
0$. Integrating over the longitudinal modes, the
Coulomb interaction becomes 
\begin{eqnarray}\label{dm}
 \tilde{S}_{C}= \sum_{q}\left[
|\rho_{q}|^{2}
\frac{e^{2}}{2\varepsilon_{0}\varepsilon(q)\vec{q}^{2}} 
+\frac{|\vec{P}^{T} (q) |^{2} }{2D_{T} (q)} 
\right].
\end{eqnarray}
Here, for $|\vec{q}|\ll q_D$
\begin{equation}\label{dielectric}
\varepsilon (\bq ,i\nu_n)= 
\varepsilon_{1}+ \frac{\Omega_{0}^{2}}{\nu_n^2+ c_s^2 \vec{q}^2+\omega_{T}^2}
\end{equation}
is the renormalized
dielectric constant,  
$P_{a}^{T} (q)= (\delta_{ab} - \hat q_{a}\hat q_{b})P_{b} (q)$
are the transverse components of the polarization.
Most importantly, in action \eqref{dm} 
the quantum
critical transverse  polar modes are entirely decoupled from the
electronic degrees of freedom, exemplifying
the dark matter analogy.

Normally, low carrier density metals are considered {\sl strongly interacting},
for the ratio of Coulomb to kinetic energy, determined by 
$r_{s}=1/ (k_{F}a_{B})$, where $k_{F}\sim n_{e}^{1/3}$ is the Fermi momentum
and $a_{B}  = \frac{4 \pi \varepsilon \hbar^2}{m^* e^2}$ the Bohr radius, is very large at low densities. However, in a quantum critical polar metal, the large
upward renormalization of the dielectric constant, Eq. \eqref{dielectric}, can severely suppresses the interaction between the electrons. Indeed, the dielectric constant at the relevant electronic scales at low densities is $\epsilon\sim \epsilon(\vec{q},\omega)\vert_{q=2k_{F},\omega=\epsilon_{F}}\approx \frac{\Omega_0^2}{(2 c_s k_F)^2}\gg1$ at the polar QCP, leading to $r_s\ll1$.
Furthermore, the electronic corrections to the dielectric constant, given in RPA by
$\delta\epsilon_{RPA}=\frac{e^2}{q^2\varepsilon_{0}} \Pi_{e}({\bf q}, \nu_n)$, where $\Pi_{e} (q)$ is the dynamical susceptibility (Lindhardt function)  of the electron gas, can be neglected. Indeed, $\frac{\delta\epsilon_{RPA}}{\varepsilon}\vert_{q=2k_{F},\omega=\epsilon_{F}}\sim r_s \ll1$.

The regime considered here is in stark contrast to the conventional case of the relevant frequency being of the order of $\omega_L\ll E_F$. There, the 
strong frequency-dependence of $\epsilon$ leads to a Bardeen-Pines attraction near the Fermi surface. Additionally, at low momentum transfers $v_F q\ll \omega$, a new scale appears in the problem
- the electronic plasma frequency; its contribution to pairing is however suppressed by the factor $\varepsilon^{-1}$ \cite{ruhman2016}. Thus, in what follows we neglect this possibility, 
approximating the electron dynamical susceptibility by its long wavelength, low frequency
limit as in \eqref{dielectric}.

We next consider the effect of turning on the coupling to energy fluctuations 
in \eqref{eq:h2ph}. The presence of a finite electron density $n_e = \langle \rho_e (x) \rangle$ leads to a shift in the phonon frequency:
\begin{equation}\label{eq:deltaomT}
\omega_{L,T}^2(n_e) = \omega_{L,T}^2+2g n_{e}\varepsilon_{0}\Omega_0^2,
\end{equation}
which naturally explains the suppression of the polar state by charge doping, universally observed in polar metals \cite{kolodiazhnyi2010,tsymbal.2012,rischau.2017,Wang2019}.

The coupling of the energy fluctuations to the electron 
density fluctuations $\delta \rho_{e} (x) = \rho_{e} (x) - n_{e}$, cannot be integrated out exactly. Interactions with critical fluctuations near QCPs can be relevant perturbations in scaling sense \cite{sachdev.1999}, destabilizing the Fermi liquid ground state already at weak coupling \cite{sachdev.1999,coleman2001}. In our case, however, the interaction Eq. \eqref{eq:h2ph} preserves the Fermi liquid. Assuming the dynamical critical exponent $z=1$ and taking the scaling dimension of momentum $[q] = 1$, one obtains  $[g] = 2 - d$, irrelevant in 3D \cite{Note1}, implying that the system remains a Fermi liquid even at the QCP. Thus, we can consider its effects perturbatively for weak coupling. Integrating out the field $\vec{P}(x)$ to lowest order in $g$, we obtain an effective interaction between electrons:
\begin{equation}\label{deltaS}
\Delta S = \frac{1}{2}\int d^{4}x d^4x' \delta \rho_{e} (x) V_{En}
(x-x')
\delta \rho_{e} (x')+ O (g^{3})
\end{equation}
where 
\begin{equation}\label{eq:ven}
V_{En} (x-x') = -2 g^{2 }{\rm Tr}\biggl[D (x-x')^{2}\biggr]
\end{equation}
is recognized to be an attractive density-density interaction resulting from two-phonon exchange, Fig. \ref{fig:cartoon} (b). At criticality, the contribution to Eq. \eqref{eq:ven} of the transverse modes stems from their propagator
\begin{equation}\label{eq:d_tr}
D_{ab}^{tr} (\vec{x},\tau )=\varepsilon_{0}
\left(\frac{\Omega_{0}}{2\pi c_{s}}\right)^{2}
\frac{1}{\vec{x}^{2}+c_{s}^{2}\tau^{2}} (\delta_{ab}-\hat x_{a}\hat  x_{b}),
\end{equation}
leading to a long-range interaction of the form
\begin{equation}\label{eq:vcrit}
V (\vec{x},\tau )= 
- \frac{\lambda^{2}}{( \vec{x}^{2}+c_{s}^{2}\tau^{2})^{2}}, 
\end{equation}
where $\lambda = \frac{g\epsilon_{0}\Omega_{0}^{2}}{2 \pi^{2}c_{s}}$. Away from criticality, \eqref{eq:vcrit} is valid for space-time separations smaller than the correlation length $\xi =c_{s}/\omega_{T}$. The interaction at finite momentum and frequency transfer, relevant for pairing, is obtained by a Fourier transform of this expression
\begin{eqnarray}\label{l}
V(i\omega,\bq )&= & 
- 
\left(\frac{\lambda}{c_{s}} \right)^{2}\int^{\xi}_{a_{0}}
\frac{e^{i[(\vec{q}\cdot\vec{x})+\omega\tau]}}{x^4}d^4x\cr
&\sim&- 
\left(\frac{2\pi^{2}\lambda^{2}}{c_{s}} \right)
\ln 
\left[
\frac{a_{0}^{-1}}
{{\rm max}(\xi^{-1}, \omega_{\bf q}/c_{s}) }
\right].
\end{eqnarray}
where $\omega_{q}= \sqrt{\omega^{2}+c_{s}^{2}\vec{q}^{2}}$. The
characteristic electronic momentum and energy transfer scales are
$q\sim k_F~\sim n^{1/3}$ and $\omega\sim E_F\sim n^{2/3}$,
respectively, resulting in $|q|^{-1}\sim n^{-1/3},\;c_s/|\omega_n|\sim
n^{-2/3}$. Furthermore, following \eqref{eq:deltaomT}, a finite 
electron density
leads to a finite correlation length of the order $\xi\sim n^{-1/2}$.

\begin{figure}[h]
	\includegraphics[width=0.4\textwidth]{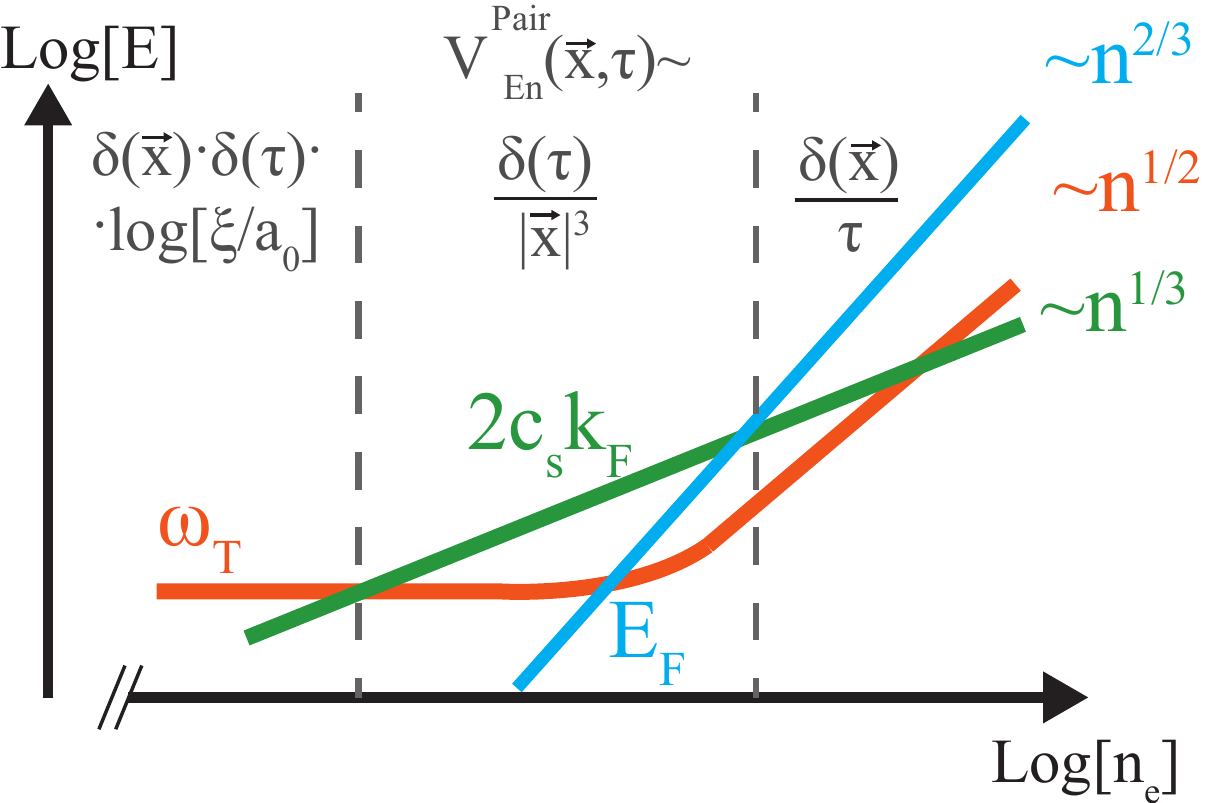}
	\caption{Density dependence of the effective energy-fluctuation mediated electron-electron attraction $V_{En}^{Pair}(\vec{x},\tau)$ (see text): colored lines show the relevant energy scales for the interaction; in each region (gray dashed lines) the dominant scale determines the effective form of the interaction. At low densities the interaction can be approximated with a local one. On increasing the density, the momentum dependence of the interaction first becomes prominent, while at the highest densities, strong retardation shows on the scales of order $\hbar/E_F$.}
	\label{fig:Enscales}
\end{figure}

Consequently, the interaction character changes with density, and
 can be described by an effective interaction
 $V_{En}^{Pair}(\vec{x},\tau)$ obtained by inverse Fourier
 transforming the final result of Eq. \eqref{l}
 (Fig. \ref{fig:Enscales}). For a polar metal critical at zero doping ($\omega_T(n_e=0)= 0$),
 one expects that at low densities the interaction is cut off at
 the momentum scale of $k_F$, while being essentially 
independent of the frequency transferred, since $E_F/c_s\ll k_F$. This is
 identical to an instantaneous repulsion, nonlocal in real space
 (Fig. \ref{fig:Enscales}, middle). In the high density
 limit, the frequency-dependence of the interaction can be no longer ignored and is suppressed for frequencies beyond $c_s k_F$, qualitatively similar to a usual phonon-mediated attraction (Fig. \ref{fig:Enscales}, rightmost region). 
Finally, if $\omega_T\gg E_F, c_s k_F$ (leftmost region of Fig. 
\ref{fig:Enscales}), 
which can be realized away from QCP or at intermediate densities, the interaction $V_{En}^{Pair}(\vec{x},\tau)$ reduces to an instantaneous local attraction.

A more detailed calculation of the interaction potential yields
\begin{equation}
\begin{gathered}
V_{2TO}(i\omega,{\bf q})
=- \left(\frac{2\pi^{2}\lambda^{2}}{c_{s}} \right)
\left(
\log \left[ \frac{\Omega_T}{\omega_{T}} \right]  -
f\biggl (\frac{\omega_q }{\omega_{T}}
\biggr)
\right),
\\
f(x) = \frac{\sqrt{4+x^{2}}}{2x}
\log \left[ \frac{\sqrt{4+x^{2}}+x}{\sqrt{4+x^2}-x} \right]-\frac{1}{2}
\label{eq:2phint},
\end{gathered}
\end{equation}
where $\Omega_T = \max_{\vec{q}} \omega_T (\vec{q})$.
Here it is important to
note, that since the integral is logarithmically divergent, large
momenta of the order of the Brillouin zone size contribute
significantly. At such momenta, the dispersion is expected to deviate
from the simple quadratic one. Thus in practical calculations, 
we need an average value of $c_s$ which we approximate in the spirit of Debye approximation by $\overline{c}_s = \Omega_T/((6\pi^2)^{1/3}/a_0)$. We remark that the contribution of the longitudinal modes to the induced interaction can be shown to be negligible in the critical regime $\Omega_T\gg\omega_{T},c_sk_F$ and assuming $\frac{\Omega_0^2}{c_s^2\kappa_D^2}\gg1$ \cite{Note1}. Additionally, in deriving the above we assumed that the momentum and energy cutoff scales are sufficiently large such that the corrections to the phonon propagator Eq. \eqref{eq:d_tr} due to phonon-phonon interactions (logarithmic for a QCP in 3D \cite{roussev.2003}), can be neglected.

{\it Superconductivity at low densities:} We now show that the attractive interaction mediated by the energy fluctuations {\it always} leads to superconductivity at low densities close to the 
polar QCP. More specifically, averaging the repulsive Coulomb 
((\ref{dm}), (\ref{dielectric})) and the attractive 
(\ref{eq:2phint}),
interactions over the Fermi surface, i.e. $\langle V(k-k') \rangle = \langle V(k_F,k_F,\theta) \rangle_\theta$, we obtain:
	\begin{equation}
		\lambda_{eff} = 
		N(0)\left(
		\left(\frac{2\pi^{2}\lambda^{2}}{c_{s}} \right)
		\log \left[ \frac{\Omega_T}{2 c_s k_F} \right]
		-
		\frac{4 \pi e^2c_s^2}
		{\Omega_0^2}
		\right)
		,
		\label{eq:lambdaeff}
		\end{equation}
	where $N(0)=\frac{k_F m^*}{2 \pi^2 \hbar^2}$ is the density of states.
	In deriving this we used that $\omega_{T} \sim (g n)^{1/2}, E_F\sim n^{2/3}\ll c_s k_F\sim n^{1/3}$. Most importantly, we find that at low enough dopings the two-phonon attraction inevitably overcomes the Coulomb repulsion due to the logarithmic enhancement of the former. In particular, the total interaction is still attractive for densities below
	\begin{equation}
	\begin{gathered}
	n<n_{cr} = \frac{1}{4a_0^3} 
	e^{-3 \alpha},
\;
	\alpha = \frac{128 \pi^5 c^5 e^2}{\Omega_0^6 g^2}
	\end{gathered}
	\end{equation}
where $\alpha\sim 40 \frac{a_B}{a_0} \frac{ E_h
\Omega_T^5}{\Omega_0^6}$ for $g\sim a_0^3$. For strongly polar
materials, $\Omega_0\gg \Omega_T$ and hence $\alpha\ll1$, so this
restriction is unimportant. Moreover, from \eqref{eq:lambdaeff} the attractive part of the
interaction behaves as $\lambda_{{eff}}\sim n^{1/3} \log (\Omega_T/c_s
n^{1/3})$, describing a dome-shaped behavior of the attractive coupling constant peaked at $k_F=\Omega_T/2 e c_s$ corresponding to a density:
	\begin{equation}\label{eq:nmax}
	n_{max} = \frac{1}{3\pi^2} \left(\frac{\Omega_T}{2 e c_s}\right)^3
\approx
	\frac{0.01 (\overline{c}_s/c_s)^3}{a_0^3}.
	\end{equation}
As the phonon dispersion flattens near the Brillouin zone edges, 
the average $\overline{c}_s<c_s$ so that $n_{max}\ll 1$ corresponds to a dilute charge concentration well below half filling. Finally, away from the QCP (i.e. $\omega_T(n_e=0)\neq0$), the Coulomb screening is reduced, resulting in an additional repulsive term $\sim 2\pi e^2 \omega_T^2/(\Omega_0^2 k_F^2)$ (c.f. \eqref{dielectric}). Due to its singular nature at $k_F\to0$ this sets a lower bound on the density $n_{min}\sim \xi^{-3} [\alpha/\log(\Omega_T/\omega_T)]^{3/2}/3\pi^2$ where the interaction is attractive.

Let us now discuss the critical temperatures of the resulting superconductor. At low densities, the interaction is essentially instantaneous (see Fig. \ref{fig:Enscales}).
The critical temperature can then be found in the non-adiabatic
weak-coupling limit to be $T_c \approx 0.28 E_F e^{-1/\lambda_{eff}}$
\cite{gorkov1961,gorkov2016prb,chubukov2016}. Due to the exponential dependence on
the coupling constant, one expects $T_c$ to have a dome-like shape
with a maximum at $n_{max}$ as in \eqref{eq:nmax}. The theory developed
here also has important consequences for the dependence of $T_c$ on
external tuning parameters (e.g. pressure)in the vicinity of a polar
QCP. Neglecting the residual Coulomb term in
\eqref{eq:lambdaeff} and thus assuming the dominance of the
energy-fluctuation attraction,  one obtains for $\omega_T\gg c_s k_F$
that $\frac{d \ln  T_c}{d \ln  \omega_T} \sim -1/  \log^2
\frac{\Omega_T}{\omega_T}$, leading to a singular dependence of
$T_{c}$ on tuning parameter near the QCP. However, in the high-density limit $2c_sk_F\ll \omega_T$, the
coupling constant is almost independent of the TO phonon frequency, so the tuning sensitivity will be much weaker, $\frac{d \log  T_c}{d \log\omega_T} \sim 
\log \left(\frac{c_s k_F}{\omega_T}\right)\omega_T^2/(c_s k_F)^2$.

{\it 2D polar metals:} Similar calculations can be performed in two dimensions. While at tree level, Eq. \eqref{eq:h2ph} is marginal, the corrections due to quartic interactions between phonons reduce \cite{kleinert2003} the momentum-space singularities of the energy fluctuations, preserving the Fermi liquid in 2D. The 2D Fourier transformation of expression \eqref{l} yields 
\begin{equation}\label{}
V_{2TO}^{2D}(i\omega_n,\vec{q})\sim \frac{g_{2D}^2\Omega_0^4}{(4\pi)^3 c_s^2} \frac{1}{
		{{\rm max}(\omega_T, \omega_{\bf q}) }
	} 
\end{equation}
- a stronger singularity than in 3D. In the limit $c_s k_F\gg
\omega_T$ the terms due to LO phonon energy fluctuation also have
to be included, being of the same order, but this does not change the
qualitative form of the interaction\cite{Note1}. Finally, the bare Coulomb repulsion in 2D is given by $\frac{2\pi e^2}{q}$. Screening with the polar mode and conduction electrons, however, reduces it to
	\begin{equation}
	V_C^{2D} (i\omega,{\bf q}) = 
	\frac{4\pi}
	{\frac{\Omega_0^2 l_0 {\bf q}^2}{\omega_n^2+\omega_T^2+c^2{\bf q}^2}+4 \frac{m^* e^2}{\hbar^2}},
	\end{equation}
	where $l_0$ is the 2D layer thickness.

\begin{figure}[h]
	\includegraphics[width=0.5\textwidth]{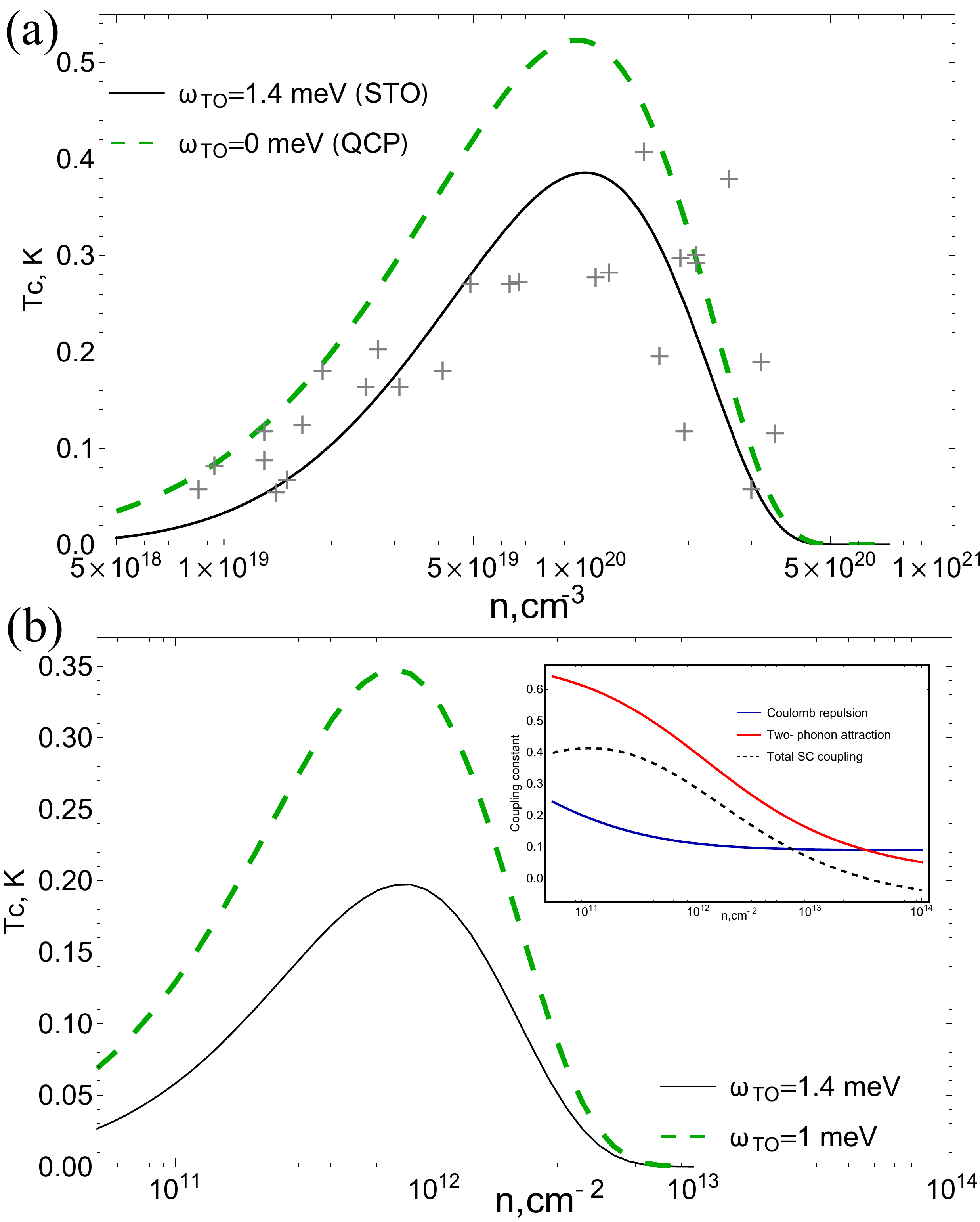}
	\caption{ (a) $T_c$ as a function of carrier density for parameters appropriate for SrTiO$_3$, where the best fit is obtained for $g/a_0^3=0.72$. The experimental  $T_c$ is determined from the onset of the Meissner effect\cite{koonce1967,collignon2017}. The green line shows the $T_c(n)$ expected at the QCP.
		(b) same as (a) for a 2D system modeling a film of SrTiO$_3$ with thickness $2 a_0$. $m^*=2m_e$ is taken due to low densities. The enhancement due to proximity to the QCP (green line) is much stronger, than in 3D case (a).}
	\label{fig:Tc}
\end{figure}

{\it Superconductivity in $SrTiO_3$:} We now apply these results to doped $SrTiO_3$. Fig. \ref{fig:Tc} (a) displays the doping dependence of $T_c$
calculated using the parameters from literature \cite{Note1} and
taking $g/a_0^3 = 0.72$, a coupling constant in accord with fits to
the low temperature $T^{2}$ resistivity\cite{maslov2020}. For the
energy fluctuation interaction, Eq.\eqref{eq:2phint}, we assumed $2
c_sk_F\gg E_F$, which holds in $SrTiO_3$ for densities lower than $2.6 \cdot
10^{19}cm^{-3}$. However, even at largest densities considered, $E_F\sim 3 (2 c_s k_F)$; as the frequency-dependence of the interaction does not alter the leading logarithmic contribution in the low-temperature limit of the pairing problem \cite{Note1}, we expect our approach to be qualitatively correct for all the relevant densities. In this approximation, 
$2\Delta/T_c=3.53$ takes the BCS value
\cite{gorkov1961,gorkov2016prb}, in accord with STM experiments
\cite{swartz2018}. Both the magnitude and the doping dependence of the critical
temperature are in good agreement 
with experiment. The dome-like shape of $T_c(n)$ arises from the 
nonomonotonic dependence of the two-phonon attractive coupling
constant on density, initially rising with the density of
states, subsequently decreasing as the momentum cutoff $c_s k_F$
approaches $\Omega_T$ (with the maximal value expected from
\eqref{eq:2phint} reached at $n_{max}$, Eq. \eqref{eq:nmax}). This is
further corroboration of the competition with Coulomb repulsion, which
becomes less screened as $q\sim k_F$ grows. We note that experiments which observe a second $T_{c}$ dome at lower doping
also suggest that superconductivity at low doping is
highly inhomogeneous and affected by the nature of the dopants
\cite{gastiasoro2020rev}, effects that lie beyond the current model.

As discussed above, proximity to the QCP should enhance $T_c$, particularly at low densities. Such a correlation has been observed for the cases of oxygen isotope substitution \cite{stucky2016} as well as Ca-Sr substitution \cite{rischau.2017}, pressure \cite{enderlein2020} or strain \cite{ahadi2019,stemmer2019,sochnikov2019}. In particular, the enhancement is observed to be more pronounced at low dopings \cite{rischau.2017}, in qualitative agreement with the arguments given above. Note also, that in the polar phase away from QCP the interaction \eqref{eq:h2ph} would still lead to pairing, however the mode frequency would grow faster than on the disordered side, according to Landau theory. Finally, \cite{enderlein2020} in a recent experiment both $T_c$ and $\varepsilon_0$ have been measured as a function of pressure. With the coupling constant being $\lambda\approx0.25$ from the known $E_F$, one obtains \cite{Note1} $\frac{d T_c}{d x} \approx 0.1 K/kbar$ consistent with the experimental value $0.06 K/kbar$ \cite{enderlein2020}.

Thin films of SrTiO$_3$ will  provide a platform to explore the
predictions of the energy fluctuation theory. Fig. \ref{fig:Tc} (b) displays the predicted $T_{c}$, using $T_c
= 0.15 E_F e^{-1/\lambda}$ for 2D instantaneous pairing ($c_s k_F\gg
\omega_T$ ) \cite{gorkov1961,gorkov2016prb,chubukov2016}), using
 coupling constants appropriate for a two-layer thick slab of SrTiO$_3$.
Assuming that the electons and phonons are in the lowest lateral quantization state, we obtain $g_{2D} = g/(2a_0)$ \cite{Note1}. The results show that an appreciable $T_c$ is obtained at low densities. Furthermore, $T_c$ is highly sensitive to the approach to criticality: a slight decrease of the TO phonon frequency essentially doubles $T_c$.

\noindent {\it Conclusion:} Here we have presented a new mechanism for 
superconductivity in quantum critical polar metals that relies neither
on retardation nor on momentum-dependence:  the electrons interact with 
quantum critical energy fluctuations of the order parameter
similar to the gravitational interactions of baryons with the dark matter. We show that this coupling of the electron density to critical energy fluctuations results naturally in a dome-shaped dependence of $T_c$ on the carrier density $n$; our estimates show that this $T_c(n)$ is consistent with that observed in doped superconducting SrTiO$_3$.
We predict that in 2D systems, e.g. epitaxial SrTiO$_3$ films, 
the effects of energy fluctuations will be even more marked, 
with extreme sensitivity of $T_c$ to the vicinity to the QCP.
From a broader perspective, large energy fluctuations have been
shown to exist near quantum critical points in the
strong-coupling regime \cite{abrahams2011,abrahams2014}; our work
presents a new mechanism for superconductivity in such strongly
correlated systems. The occurence of superconductivity mediated by energy-density fluctuations can also serve as a tool to probe the "dark matter" aspects of the solid state, involving excitations that do not interact electromagnetically, such as those related to hidden orders.

\noindent {\it Acknowledgments:} The authors gratefully 
acknowledge stimulating discussions with D. L. Maslov, and with the late
P. W. Anderson during early stages of this work.  P. A. V. is  supported 
by a Rutgers Center for Materials Theory Fellowship, P. Chandra is  
supported by DOE Basic Energy Sciences grant DE-SC0020353 and 
P. Coleman is supported by  NSF grant DMR-1830707.

\bibliography{refs_sc}

\begin{thebibliography}{48}%
\makeatletter
\providecommand \@ifxundefined [1]{%
 \@ifx{#1\undefined}
}%
\providecommand \@ifnum [1]{%
 \ifnum #1\expandafter \@firstoftwo
 \else \expandafter \@secondoftwo
 \fi
}%
\providecommand \@ifx [1]{%
 \ifx #1\expandafter \@firstoftwo
 \else \expandafter \@secondoftwo
 \fi
}%
\providecommand \natexlab [1]{#1}%
\providecommand \enquote  [1]{``#1''}%
\providecommand \bibnamefont  [1]{#1}%
\providecommand \bibfnamefont [1]{#1}%
\providecommand \citenamefont [1]{#1}%
\providecommand \href@noop [0]{\@secondoftwo}%
\providecommand \href [0]{\begingroup \@sanitize@url \@href}%
\providecommand \@href[1]{\@@startlink{#1}\@@href}%
\providecommand \@@href[1]{\endgroup#1\@@endlink}%
\providecommand \@sanitize@url [0]{\catcode `\\12\catcode `\$12\catcode
  `\&12\catcode `\#12\catcode `\^12\catcode `\_12\catcode `\%12\relax}%
\providecommand \@@startlink[1]{}%
\providecommand \@@endlink[0]{}%
\providecommand \url  [0]{\begingroup\@sanitize@url \@url }%
\providecommand \@url [1]{\endgroup\@href {#1}{\urlprefix }}%
\providecommand \urlprefix  [0]{URL }%
\providecommand \Eprint [0]{\href }%
\providecommand \doibase [0]{http://dx.doi.org/}%
\providecommand \selectlanguage [0]{\@gobble}%
\providecommand \bibinfo  [0]{\@secondoftwo}%
\providecommand \bibfield  [0]{\@secondoftwo}%
\providecommand \translation [1]{[#1]}%
\providecommand \BibitemOpen [0]{}%
\providecommand \bibitemStop [0]{}%
\providecommand \bibitemNoStop [0]{.\EOS\space}%
\providecommand \EOS [0]{\spacefactor3000\relax}%
\providecommand \BibitemShut  [1]{\csname bibitem#1\endcsname}%
\let\auto@bib@innerbib\@empty
\bibitem [{\citenamefont {Cooper}(1956)}]{cooper1956}%
  \BibitemOpen
  \bibfield  {author} {\bibinfo {author} {\bibfnamefont {Leon~N.}\ \bibnamefont
  {Cooper}},\ }\bibfield  {title} {\enquote {\bibinfo {title} {{Bound Electron
  Pairs in a Degenerate Fermi Gas}},}\ }\href {\doibase
  10.1103/PhysRev.104.1189} {\bibfield  {journal} {\bibinfo  {journal} {Phys.
  Rev.}\ }\textbf {\bibinfo {volume} {104}},\ \bibinfo {pages} {1189--1190}
  (\bibinfo {year} {1956})}\BibitemShut {NoStop}%
\bibitem [{\citenamefont {Bogoljubov}\ \emph {et~al.}(1958)\citenamefont
  {Bogoljubov}, \citenamefont {Tolmachov},\ and\ \citenamefont
  {Širkov}}]{tolmachev1958}%
  \BibitemOpen
  \bibfield  {author} {\bibinfo {author} {\bibfnamefont {N.~N.}\ \bibnamefont
  {Bogoljubov}}, \bibinfo {author} {\bibfnamefont {V.~V.}\ \bibnamefont
  {Tolmachov}}, \ and\ \bibinfo {author} {\bibfnamefont {D.~V.}\ \bibnamefont
  {Širkov}},\ }\bibfield  {title} {\enquote {\bibinfo {title} {A new method in
  the theory of superconductivity},}\ }\href {\doibase
  10.1002/prop.19580061102} {\bibfield  {journal} {\bibinfo  {journal}
  {Fortschritte der Physik}\ }\textbf {\bibinfo {volume} {6}},\ \bibinfo
  {pages} {605--682} (\bibinfo {year} {1958})},\ \Eprint
  {http://arxiv.org/abs/https://onlinelibrary.wiley.com/doi/pdf/10.1002/prop.19580061102}
  {https://onlinelibrary.wiley.com/doi/pdf/10.1002/prop.19580061102}
  \BibitemShut {NoStop}%
\bibitem [{\citenamefont {Morel}\ and\ \citenamefont
  {Anderson}(1962)}]{morel1962}%
  \BibitemOpen
  \bibfield  {author} {\bibinfo {author} {\bibfnamefont {P.}~\bibnamefont
  {Morel}}\ and\ \bibinfo {author} {\bibfnamefont {P.~W.}\ \bibnamefont
  {Anderson}},\ }\bibfield  {title} {\enquote {\bibinfo {title} {Calculation of
  the superconducting state parameters with retarded electron-phonon
  interaction},}\ }\href {\doibase 10.1103/PhysRev.125.1263} {\bibfield
  {journal} {\bibinfo  {journal} {Phys. Rev.}\ }\textbf {\bibinfo {volume}
  {125}},\ \bibinfo {pages} {1263--1271} (\bibinfo {year} {1962})}\BibitemShut
  {NoStop}%
\bibitem [{\citenamefont {Gurevich}\ \emph {et~al.}(1962)\citenamefont
  {Gurevich}, \citenamefont {Larkin},\ and\ \citenamefont
  {Firsov}}]{gurevich1962}%
  \BibitemOpen
  \bibfield  {author} {\bibinfo {author} {\bibfnamefont {VL}~\bibnamefont
  {Gurevich}}, \bibinfo {author} {\bibfnamefont {AI}~\bibnamefont {Larkin}}, \
  and\ \bibinfo {author} {\bibfnamefont {Yu~A}\ \bibnamefont {Firsov}},\
  }\bibfield  {title} {\enquote {\bibinfo {title} {Possibility of
  superconductivity in semiconductors},}\ }\href@noop {} {\bibfield  {journal}
  {\bibinfo  {journal} {Sov. Phys.-Solid State (Engl. Transl.);(United
  States)}\ }\textbf {\bibinfo {volume} {4}} (\bibinfo {year}
  {1962})}\BibitemShut {NoStop}%
\bibitem [{\citenamefont {Collignon}\ \emph {et~al.}(2019)\citenamefont
  {Collignon}, \citenamefont {Lin}, \citenamefont {Rischau}, \citenamefont
  {Fauque},\ and\ \citenamefont {Behnia}}]{collignon2019}%
  \BibitemOpen
  \bibfield  {author} {\bibinfo {author} {\bibfnamefont {Clement}\ \bibnamefont
  {Collignon}}, \bibinfo {author} {\bibfnamefont {Xiao}\ \bibnamefont {Lin}},
  \bibinfo {author} {\bibfnamefont {Carl~Willem}\ \bibnamefont {Rischau}},
  \bibinfo {author} {\bibfnamefont {Benoit}\ \bibnamefont {Fauque}}, \ and\
  \bibinfo {author} {\bibfnamefont {Kamran}\ \bibnamefont {Behnia}},\
  }\bibfield  {title} {\enquote {\bibinfo {title} {Metallicity and
  superconductivity in doped strontium titanate},}\ }\href {\doibase
  10.1146/annurev-conmatphys-031218-013144} {\bibfield  {journal} {\bibinfo
  {journal} {Annual Review of Condensed Matter Physics}\ }\textbf {\bibinfo
  {volume} {10}},\ \bibinfo {pages} {25--44} (\bibinfo {year} {2019})},\
  \Eprint
  {http://arxiv.org/abs/https://doi.org/10.1146/annurev-conmatphys-031218-013144}
  {https://doi.org/10.1146/annurev-conmatphys-031218-013144} \BibitemShut
  {NoStop}%
\bibitem [{\citenamefont {Lin}\ \emph {et~al.}(2015)\citenamefont {Lin},
  \citenamefont {Rischau}, \citenamefont {van~der Beek}, \citenamefont
  {Fauqu\'e},\ and\ \citenamefont {Behnia}}]{lin2015}%
  \BibitemOpen
  \bibfield  {author} {\bibinfo {author} {\bibfnamefont {Xiao}\ \bibnamefont
  {Lin}}, \bibinfo {author} {\bibfnamefont {Carl~Willem}\ \bibnamefont
  {Rischau}}, \bibinfo {author} {\bibfnamefont {Cornelis~J.}\ \bibnamefont
  {van~der Beek}}, \bibinfo {author} {\bibfnamefont {Beno\^{\i}t}\ \bibnamefont
  {Fauqu\'e}}, \ and\ \bibinfo {author} {\bibfnamefont {Kamran}\ \bibnamefont
  {Behnia}},\ }\bibfield  {title} {\enquote {\bibinfo {title} {{$s$-wave
  superconductivity in optimally doped
  ${\mathrm{SrTi}}_{1\ensuremath{-}x}{\mathrm{Nb}}_{x}{\mathrm{O}}_{3}$
  unveiled by electron irradiation}},}\ }\href {\doibase
  10.1103/PhysRevB.92.174504} {\bibfield  {journal} {\bibinfo  {journal} {Phys.
  Rev. B}\ }\textbf {\bibinfo {volume} {92}},\ \bibinfo {pages} {174504}
  (\bibinfo {year} {2015})}\BibitemShut {NoStop}%
\bibitem [{\citenamefont {Swartz}\ \emph {et~al.}(2018)\citenamefont {Swartz},
  \citenamefont {Inoue}, \citenamefont {Merz}, \citenamefont {Hikita},
  \citenamefont {Raghu}, \citenamefont {Devereaux}, \citenamefont {Johnston},\
  and\ \citenamefont {Hwang}}]{swartz2018}%
  \BibitemOpen
  \bibfield  {author} {\bibinfo {author} {\bibfnamefont {Adrian~G.}\
  \bibnamefont {Swartz}}, \bibinfo {author} {\bibfnamefont {Hisashi}\
  \bibnamefont {Inoue}}, \bibinfo {author} {\bibfnamefont {Tyler~A.}\
  \bibnamefont {Merz}}, \bibinfo {author} {\bibfnamefont {Yasuyuki}\
  \bibnamefont {Hikita}}, \bibinfo {author} {\bibfnamefont {Srinivas}\
  \bibnamefont {Raghu}}, \bibinfo {author} {\bibfnamefont {Thomas~P.}\
  \bibnamefont {Devereaux}}, \bibinfo {author} {\bibfnamefont {Steven}\
  \bibnamefont {Johnston}}, \ and\ \bibinfo {author} {\bibfnamefont
  {Harold~Y.}\ \bibnamefont {Hwang}},\ }\bibfield  {title} {\enquote {\bibinfo
  {title} {Polaronic behavior in a weak-coupling superconductor},}\ }\href
  {\doibase 10.1073/pnas.1713916115} {\bibfield  {journal} {\bibinfo  {journal}
  {Proceedings of the National Academy of Sciences}\ }\textbf {\bibinfo
  {volume} {115}},\ \bibinfo {pages} {1475--1480} (\bibinfo {year} {2018})},\
  \Eprint
  {http://arxiv.org/abs/https://www.pnas.org/content/115/7/1475.full.pdf}
  {https://www.pnas.org/content/115/7/1475.full.pdf} \BibitemShut {NoStop}%
\bibitem [{\citenamefont {Gor{\textquoteright}kov}(2016)}]{gorkovpnas}%
  \BibitemOpen
  \bibfield  {author} {\bibinfo {author} {\bibfnamefont {Lev~P.}\ \bibnamefont
  {Gor{\textquoteright}kov}},\ }\bibfield  {title} {\enquote {\bibinfo {title}
  {{Phonon mechanism in the most dilute superconductor n-type SrTiO$_3$}},}\
  }\href {\doibase 10.1073/pnas.1604145113} {\bibfield  {journal} {\bibinfo
  {journal} {Proceedings of the National Academy of Sciences}\ }\textbf
  {\bibinfo {volume} {113}},\ \bibinfo {pages} {4646--4651} (\bibinfo {year}
  {2016})},\ \Eprint
  {http://arxiv.org/abs/https://www.pnas.org/content/113/17/4646.full.pdf}
  {https://www.pnas.org/content/113/17/4646.full.pdf} \BibitemShut {NoStop}%
\bibitem [{\citenamefont {Takada}(1978)}]{takada1978}%
  \BibitemOpen
  \bibfield  {author} {\bibinfo {author} {\bibfnamefont {Yasutami}\
  \bibnamefont {Takada}},\ }\bibfield  {title} {\enquote {\bibinfo {title}
  {{Plasmon Mechanism of Superconductivity in Two- and Three-Dimensional
  Electron Systems}},}\ }\href {\doibase 10.1143/JPSJ.45.786} {\bibfield
  {journal} {\bibinfo  {journal} {Journal of the Physical Society of Japan}\
  }\textbf {\bibinfo {volume} {45}},\ \bibinfo {pages} {786--794} (\bibinfo
  {year} {1978})}\BibitemShut {NoStop}%
\bibitem [{\citenamefont {Takada}(1980)}]{takada1980}%
  \BibitemOpen
  \bibfield  {author} {\bibinfo {author} {\bibfnamefont {Yasutami}\
  \bibnamefont {Takada}},\ }\bibfield  {title} {\enquote {\bibinfo {title}
  {{Theory of Superconductivity in Polar Semiconductors and Its Application to
  N-Type Semiconducting SrTiO3}},}\ }\href {\doibase 10.1143/JPSJ.49.1267}
  {\bibfield  {journal} {\bibinfo  {journal} {Journal of the Physical Society
  of Japan}\ }\textbf {\bibinfo {volume} {49}},\ \bibinfo {pages} {1267--1275}
  (\bibinfo {year} {1980})}\BibitemShut {NoStop}%
\bibitem [{\citenamefont {Ruhman}\ and\ \citenamefont
  {Lee}(2016)}]{ruhman2016}%
  \BibitemOpen
  \bibfield  {author} {\bibinfo {author} {\bibfnamefont {Jonathan}\
  \bibnamefont {Ruhman}}\ and\ \bibinfo {author} {\bibfnamefont {Patrick~A.}\
  \bibnamefont {Lee}},\ }\bibfield  {title} {\enquote {\bibinfo {title}
  {Superconductivity at very low density: The case of strontium titanate},}\
  }\href {\doibase 10.1103/PhysRevB.94.224515} {\bibfield  {journal} {\bibinfo
  {journal} {Phys. Rev. B}\ }\textbf {\bibinfo {volume} {94}},\ \bibinfo
  {pages} {224515} (\bibinfo {year} {2016})}\BibitemShut {NoStop}%
\bibitem [{\citenamefont {Gor'kov}(2017)}]{gorkov2017}%
  \BibitemOpen
  \bibfield  {author} {\bibinfo {author} {\bibfnamefont {Lev~P.}\ \bibnamefont
  {Gor'kov}},\ }\bibfield  {title} {\enquote {\bibinfo {title} {Back to
  mechanisms of superconductivity in low-doped strontium titanate},}\ }\href
  {\doibase 10.1007/s10948-017-4000-1} {\bibfield  {journal} {\bibinfo
  {journal} {Journal of Superconductivity and Novel Magnetism}\ }\textbf
  {\bibinfo {volume} {30}},\ \bibinfo {pages} {845--852} (\bibinfo {year}
  {2017})}\BibitemShut {NoStop}%
\bibitem [{\citenamefont {Enderlein}\ \emph {et~al.}(2020)\citenamefont
  {Enderlein}, \citenamefont {de~Oliveira}, \citenamefont {Tompsett},
  \citenamefont {Saitovitch}, \citenamefont {Saxena}, \citenamefont
  {Lonzarich},\ and\ \citenamefont {Rowley}}]{enderlein2020}%
  \BibitemOpen
  \bibfield  {author} {\bibinfo {author} {\bibfnamefont {C.}~\bibnamefont
  {Enderlein}}, \bibinfo {author} {\bibfnamefont {J.~Ferreira}\ \bibnamefont
  {de~Oliveira}}, \bibinfo {author} {\bibfnamefont {D.~A.}\ \bibnamefont
  {Tompsett}}, \bibinfo {author} {\bibfnamefont {E.~Baggio}\ \bibnamefont
  {Saitovitch}}, \bibinfo {author} {\bibfnamefont {S.~S.}\ \bibnamefont
  {Saxena}}, \bibinfo {author} {\bibfnamefont {G.~G.}\ \bibnamefont
  {Lonzarich}}, \ and\ \bibinfo {author} {\bibfnamefont {S.~E.}\ \bibnamefont
  {Rowley}},\ }\bibfield  {title} {\enquote {\bibinfo {title}
  {Superconductivity mediated by polar modes in ferroelectric metals},}\ }\href
  {\doibase 10.1038/s41467-020-18438-0} {\bibfield  {journal} {\bibinfo
  {journal} {Nature Communications}\ }\textbf {\bibinfo {volume} {11}},\
  \bibinfo {pages} {4852} (\bibinfo {year} {2020})}\BibitemShut {NoStop}%
\bibitem [{\citenamefont {Ma}\ \emph {et~al.}(2021)\citenamefont {Ma},
  \citenamefont {Yang},\ and\ \citenamefont {Chen}}]{Ma2021}%
  \BibitemOpen
  \bibfield  {author} {\bibinfo {author} {\bibfnamefont {Jiaji}\ \bibnamefont
  {Ma}}, \bibinfo {author} {\bibfnamefont {Ruihan}\ \bibnamefont {Yang}}, \
  and\ \bibinfo {author} {\bibfnamefont {Hanghui}\ \bibnamefont {Chen}},\
  }\bibfield  {title} {\enquote {\bibinfo {title} {A large modulation of
  electron-phonon coupling and an emergent superconducting dome in doped strong
  ferroelectrics},}\ }\href {\doibase 10.1038/s41467-021-22541-1} {\bibfield
  {journal} {\bibinfo  {journal} {Nature Communications}\ }\textbf {\bibinfo
  {volume} {12}},\ \bibinfo {pages} {2314} (\bibinfo {year}
  {2021})}\BibitemShut {NoStop}%
\bibitem [{\citenamefont {Kanasugi}\ and\ \citenamefont
  {Yanase}(2018)}]{yanase2018}%
  \BibitemOpen
  \bibfield  {author} {\bibinfo {author} {\bibfnamefont {Shota}\ \bibnamefont
  {Kanasugi}}\ and\ \bibinfo {author} {\bibfnamefont {Youichi}\ \bibnamefont
  {Yanase}},\ }\bibfield  {title} {\enquote {\bibinfo {title}
  {Spin-orbit-coupled ferroelectric superconductivity},}\ }\href {\doibase
  10.1103/PhysRevB.98.024521} {\bibfield  {journal} {\bibinfo  {journal} {Phys.
  Rev. B}\ }\textbf {\bibinfo {volume} {98}},\ \bibinfo {pages} {024521}
  (\bibinfo {year} {2018})}\BibitemShut {NoStop}%
\bibitem [{\citenamefont {Gastiasoro}\ \emph
  {et~al.}(2020{\natexlab{a}})\citenamefont {Gastiasoro}, \citenamefont
  {Trevisan},\ and\ \citenamefont {Fernandes}}]{gastiasoro2020}%
  \BibitemOpen
  \bibfield  {author} {\bibinfo {author} {\bibfnamefont {Maria~N.}\
  \bibnamefont {Gastiasoro}}, \bibinfo {author} {\bibfnamefont {Tha\'{\i}s~V.}\
  \bibnamefont {Trevisan}}, \ and\ \bibinfo {author} {\bibfnamefont
  {Rafael~M.}\ \bibnamefont {Fernandes}},\ }\bibfield  {title} {\enquote
  {\bibinfo {title} {{Anisotropic superconductivity mediated by ferroelectric
  fluctuations in cubic systems with spin-orbit coupling}},}\ }\href {\doibase
  10.1103/PhysRevB.101.174501} {\bibfield  {journal} {\bibinfo  {journal}
  {Phys. Rev. B}\ }\textbf {\bibinfo {volume} {101}},\ \bibinfo {pages}
  {174501} (\bibinfo {year} {2020}{\natexlab{a}})}\BibitemShut {NoStop}%
\bibitem [{\citenamefont {Kanasugi}\ \emph {et~al.}(2020)\citenamefont
  {Kanasugi}, \citenamefont {Kuzmanovski}, \citenamefont {Balatsky},\ and\
  \citenamefont {Yanase}}]{kanasugi2020}%
  \BibitemOpen
  \bibfield  {author} {\bibinfo {author} {\bibfnamefont {Shota}\ \bibnamefont
  {Kanasugi}}, \bibinfo {author} {\bibfnamefont {Dushko}\ \bibnamefont
  {Kuzmanovski}}, \bibinfo {author} {\bibfnamefont {Alexander~V.}\ \bibnamefont
  {Balatsky}}, \ and\ \bibinfo {author} {\bibfnamefont {Youichi}\ \bibnamefont
  {Yanase}},\ }\bibfield  {title} {\enquote {\bibinfo {title}
  {{Ferroelectricity-induced multiorbital odd-frequency superconductivity in
  $\mathrm{Sr}\mathrm{Ti}{\mathrm{O}}_{3}$}},}\ }\href {\doibase
  10.1103/PhysRevB.102.184506} {\bibfield  {journal} {\bibinfo  {journal}
  {Phys. Rev. B}\ }\textbf {\bibinfo {volume} {102}},\ \bibinfo {pages}
  {184506} (\bibinfo {year} {2020})}\BibitemShut {NoStop}%
\bibitem [{\citenamefont {Volkov}\ and\ \citenamefont
  {Chandra}(2020)}]{volkov2020}%
  \BibitemOpen
  \bibfield  {author} {\bibinfo {author} {\bibfnamefont {Pavel~A.}\
  \bibnamefont {Volkov}}\ and\ \bibinfo {author} {\bibfnamefont {Premala}\
  \bibnamefont {Chandra}},\ }\bibfield  {title} {\enquote {\bibinfo {title}
  {Multiband quantum criticality of polar metals},}\ }\href {\doibase
  10.1103/PhysRevLett.124.237601} {\bibfield  {journal} {\bibinfo  {journal}
  {Phys. Rev. Lett.}\ }\textbf {\bibinfo {volume} {124}},\ \bibinfo {pages}
  {237601} (\bibinfo {year} {2020})}\BibitemShut {NoStop}%
\bibitem [{\citenamefont {Edge}\ \emph {et~al.}(2015)\citenamefont {Edge},
  \citenamefont {Kedem}, \citenamefont {Aschauer}, \citenamefont {Spaldin},\
  and\ \citenamefont {Balatsky}}]{edge.2015}%
  \BibitemOpen
  \bibfield  {author} {\bibinfo {author} {\bibfnamefont {Jonathan~M.}\
  \bibnamefont {Edge}}, \bibinfo {author} {\bibfnamefont {Yaron}\ \bibnamefont
  {Kedem}}, \bibinfo {author} {\bibfnamefont {Ulrich}\ \bibnamefont
  {Aschauer}}, \bibinfo {author} {\bibfnamefont {Nicola~A.}\ \bibnamefont
  {Spaldin}}, \ and\ \bibinfo {author} {\bibfnamefont {Alexander~V.}\
  \bibnamefont {Balatsky}},\ }\bibfield  {title} {\enquote {\bibinfo {title}
  {Quantum critical origin of the superconducting dome in
  ${\mathrm{srtio}}_{3}$},}\ }\href {\doibase 10.1103/PhysRevLett.115.247002}
  {\bibfield  {journal} {\bibinfo  {journal} {Phys. Rev. Lett.}\ }\textbf
  {\bibinfo {volume} {115}},\ \bibinfo {pages} {247002} (\bibinfo {year}
  {2015})}\BibitemShut {NoStop}%
\bibitem [{\citenamefont {Stucky}\ \emph {et~al.}(2016)\citenamefont {Stucky},
  \citenamefont {Scheerer}, \citenamefont {Ren}, \citenamefont {Jaccard},
  \citenamefont {Poumirol}, \citenamefont {Barreteau}, \citenamefont
  {Giannini},\ and\ \citenamefont {van~der Marel}}]{stucky2016}%
  \BibitemOpen
  \bibfield  {author} {\bibinfo {author} {\bibfnamefont {A.}~\bibnamefont
  {Stucky}}, \bibinfo {author} {\bibfnamefont {G.~W.}\ \bibnamefont
  {Scheerer}}, \bibinfo {author} {\bibfnamefont {Z.}~\bibnamefont {Ren}},
  \bibinfo {author} {\bibfnamefont {D.}~\bibnamefont {Jaccard}}, \bibinfo
  {author} {\bibfnamefont {J.-M.}\ \bibnamefont {Poumirol}}, \bibinfo {author}
  {\bibfnamefont {C.}~\bibnamefont {Barreteau}}, \bibinfo {author}
  {\bibfnamefont {E.}~\bibnamefont {Giannini}}, \ and\ \bibinfo {author}
  {\bibfnamefont {D.}~\bibnamefont {van~der Marel}},\ }\bibfield  {title}
  {\enquote {\bibinfo {title} {{Isotope effect in superconducting n-doped
  SrTiO$_3$}},}\ }\href {\doibase 10.1038/srep37582} {\bibfield  {journal}
  {\bibinfo  {journal} {Scientific Reports}\ }\textbf {\bibinfo {volume} {6}},\
  \bibinfo {pages} {37582} (\bibinfo {year} {2016})}\BibitemShut {NoStop}%
\bibitem [{\citenamefont {Rischau}\ \emph {et~al.}(2017)\citenamefont
  {Rischau}, \citenamefont {Lin}, \citenamefont {Grams}, \citenamefont {Finck},
  \citenamefont {Harms}, \citenamefont {Engelmayer}, \citenamefont {Lorenz},
  \citenamefont {Gallais}, \citenamefont {Fauque}, \citenamefont {Hemberger},\
  and\ \citenamefont {Behnia}}]{rischau.2017}%
  \BibitemOpen
  \bibfield  {author} {\bibinfo {author} {\bibfnamefont {Carl~Willem}\
  \bibnamefont {Rischau}}, \bibinfo {author} {\bibfnamefont {Xiao}\
  \bibnamefont {Lin}}, \bibinfo {author} {\bibfnamefont {Christoph~P.}\
  \bibnamefont {Grams}}, \bibinfo {author} {\bibfnamefont {Dennis}\
  \bibnamefont {Finck}}, \bibinfo {author} {\bibfnamefont {Steffen}\
  \bibnamefont {Harms}}, \bibinfo {author} {\bibfnamefont {Johannes}\
  \bibnamefont {Engelmayer}}, \bibinfo {author} {\bibfnamefont {Thomas}\
  \bibnamefont {Lorenz}}, \bibinfo {author} {\bibfnamefont {Yann}\ \bibnamefont
  {Gallais}}, \bibinfo {author} {\bibfnamefont {Benoit}\ \bibnamefont
  {Fauque}}, \bibinfo {author} {\bibfnamefont {Joachim}\ \bibnamefont
  {Hemberger}}, \ and\ \bibinfo {author} {\bibfnamefont {Kamran}\ \bibnamefont
  {Behnia}},\ }\bibfield  {title} {\enquote {\bibinfo {title} {{A ferroelectric
  quantum phase transition inside the superconducting dome of
  Sr$_{1-x}$Ca$_x$TiO$_{3-\delta}$}},}\ }\href@noop {} {\bibfield  {journal}
  {\bibinfo  {journal} {Nature Physics}\ }\textbf {\bibinfo {volume} {13}},\
  \bibinfo {pages} {643} (\bibinfo {year} {2017})}\BibitemShut {NoStop}%
\bibitem [{\citenamefont {Tomioka}\ \emph {et~al.}(2019)\citenamefont
  {Tomioka}, \citenamefont {Shirakawa}, \citenamefont {Shibuya},\ and\
  \citenamefont {Inoue}}]{tomioka2019}%
  \BibitemOpen
  \bibfield  {author} {\bibinfo {author} {\bibfnamefont {Yasuhide}\
  \bibnamefont {Tomioka}}, \bibinfo {author} {\bibfnamefont {Naoki}\
  \bibnamefont {Shirakawa}}, \bibinfo {author} {\bibfnamefont {Keisuke}\
  \bibnamefont {Shibuya}}, \ and\ \bibinfo {author} {\bibfnamefont {Isao~H.}\
  \bibnamefont {Inoue}},\ }\bibfield  {title} {\enquote {\bibinfo {title}
  {Enhanced superconductivity close to a non-magnetic quantum critical point in
  electron-doped strontium titanate},}\ }\href {\doibase
  10.1038/s41467-019-08693-1} {\bibfield  {journal} {\bibinfo  {journal}
  {Nature Communications}\ }\textbf {\bibinfo {volume} {10}},\ \bibinfo {pages}
  {738} (\bibinfo {year} {2019})}\BibitemShut {NoStop}%
\bibitem [{\citenamefont {W\"olfle}\ and\ \citenamefont
  {Balatsky}(2018)}]{wolfle.2018}%
  \BibitemOpen
  \bibfield  {author} {\bibinfo {author} {\bibfnamefont {Peter}\ \bibnamefont
  {W\"olfle}}\ and\ \bibinfo {author} {\bibfnamefont {Alexander~V.}\
  \bibnamefont {Balatsky}},\ }\bibfield  {title} {\enquote {\bibinfo {title}
  {{Superconductivity at low density near a ferroelectric quantum critical
  point: Doped ${\mathrm{SrTiO}}_{3}$}},}\ }\href {\doibase
  10.1103/PhysRevB.98.104505} {\bibfield  {journal} {\bibinfo  {journal} {Phys.
  Rev. B}\ }\textbf {\bibinfo {volume} {98}},\ \bibinfo {pages} {104505}
  (\bibinfo {year} {2018})}\BibitemShut {NoStop}%
\bibitem [{\citenamefont {Ruhman}\ and\ \citenamefont
  {Lee}(2019)}]{wolfle.2018com}%
  \BibitemOpen
  \bibfield  {author} {\bibinfo {author} {\bibfnamefont {Jonathan}\
  \bibnamefont {Ruhman}}\ and\ \bibinfo {author} {\bibfnamefont {Patrick~A.}\
  \bibnamefont {Lee}},\ }\bibfield  {title} {\enquote {\bibinfo {title}
  {{Comment on ``Superconductivity at low density near a ferroelectric quantum
  critical point: Doped ${\mathrm{SrTiO}}_{3}$''}},}\ }\href {\doibase
  10.1103/PhysRevB.100.226501} {\bibfield  {journal} {\bibinfo  {journal}
  {Phys. Rev. B}\ }\textbf {\bibinfo {volume} {100}},\ \bibinfo {pages}
  {226501} (\bibinfo {year} {2019})}\BibitemShut {NoStop}%
\bibitem [{\citenamefont {W\"olfle}\ and\ \citenamefont
  {Balatsky}(2019)}]{wolfle.2018repl}%
  \BibitemOpen
  \bibfield  {author} {\bibinfo {author} {\bibfnamefont {Peter}\ \bibnamefont
  {W\"olfle}}\ and\ \bibinfo {author} {\bibfnamefont {Alexander~V.}\
  \bibnamefont {Balatsky}},\ }\bibfield  {title} {\enquote {\bibinfo {title}
  {{Reply to ``Comment on `Superconductivity at low density near a
  ferroelectric quantum critical point: Doped ${\mathrm{SrTiO}}_{3}$'''}},}\
  }\href {\doibase 10.1103/PhysRevB.100.226502} {\bibfield  {journal} {\bibinfo
   {journal} {Phys. Rev. B}\ }\textbf {\bibinfo {volume} {100}},\ \bibinfo
  {pages} {226502} (\bibinfo {year} {2019})}\BibitemShut {NoStop}%
\bibitem [{\citenamefont {Ngai}(1974)}]{ngai1974}%
  \BibitemOpen
  \bibfield  {author} {\bibinfo {author} {\bibfnamefont {K.~L.}\ \bibnamefont
  {Ngai}},\ }\bibfield  {title} {\enquote {\bibinfo {title} {Two-phonon
  deformation potential and superconductivity in degenerate semiconductors},}\
  }\href {\doibase 10.1103/PhysRevLett.32.215} {\bibfield  {journal} {\bibinfo
  {journal} {Phys. Rev. Lett.}\ }\textbf {\bibinfo {volume} {32}},\ \bibinfo
  {pages} {215--218} (\bibinfo {year} {1974})}\BibitemShut {NoStop}%
\bibitem [{\citenamefont {Kumar}\ \emph {et~al.}(2021)\citenamefont {Kumar},
  \citenamefont {Yudson},\ and\ \citenamefont {Maslov}}]{maslov2020}%
  \BibitemOpen
  \bibfield  {author} {\bibinfo {author} {\bibfnamefont {Abhishek}\
  \bibnamefont {Kumar}}, \bibinfo {author} {\bibfnamefont {Vladimir~I.}\
  \bibnamefont {Yudson}}, \ and\ \bibinfo {author} {\bibfnamefont {Dmitrii~L.}\
  \bibnamefont {Maslov}},\ }\bibfield  {title} {\enquote {\bibinfo {title}
  {Quasiparticle and nonquasiparticle transport in doped quantum
  paraelectrics},}\ }\href {\doibase 10.1103/PhysRevLett.126.076601} {\bibfield
   {journal} {\bibinfo  {journal} {Phys. Rev. Lett.}\ }\textbf {\bibinfo
  {volume} {126}},\ \bibinfo {pages} {076601} (\bibinfo {year}
  {2021})}\BibitemShut {NoStop}%
\bibitem [{Note1()}]{Note1}%
  \BibitemOpen
  \bibinfo {note} {See Supplemental Material at [URL will be inserted by
  publisher] for the details.}\BibitemShut {Stop}%
\bibitem [{\citenamefont {Nazaryan}\ and\ \citenamefont
  {Feigelman}(2021)}]{nazaryan2021}%
  \BibitemOpen
  \bibfield  {author} {\bibinfo {author} {\bibfnamefont {Kh.~G.}\ \bibnamefont
  {Nazaryan}}\ and\ \bibinfo {author} {\bibfnamefont {M.~V.}\ \bibnamefont
  {Feigelman}},\ }\href@noop {} {\enquote {\bibinfo {title} {{Conductivity and
  thermoelectric coefficients of doped SrTiO$_3$ at high temperatures}},}\ }
  (\bibinfo {year} {2021}),\ \Eprint {http://arxiv.org/abs/2103.11425}
  {arXiv:2103.11425 [cond-mat.other]} \BibitemShut {NoStop}%
\bibitem [{\citenamefont {van~der Marel}\ \emph {et~al.}(2019)\citenamefont
  {van~der Marel}, \citenamefont {Barantani},\ and\ \citenamefont
  {Rischau}}]{marel2019}%
  \BibitemOpen
  \bibfield  {author} {\bibinfo {author} {\bibfnamefont {D.}~\bibnamefont
  {van~der Marel}}, \bibinfo {author} {\bibfnamefont {F.}~\bibnamefont
  {Barantani}}, \ and\ \bibinfo {author} {\bibfnamefont {C.~W.}\ \bibnamefont
  {Rischau}},\ }\bibfield  {title} {\enquote {\bibinfo {title} {{Possible
  mechanism for superconductivity in doped ${\mathrm{SrTiO}}_{3}$}},}\ }\href
  {\doibase 10.1103/PhysRevResearch.1.013003} {\bibfield  {journal} {\bibinfo
  {journal} {Phys. Rev. Research}\ }\textbf {\bibinfo {volume} {1}},\ \bibinfo
  {pages} {013003} (\bibinfo {year} {2019})}\BibitemShut {NoStop}%
\bibitem [{\citenamefont {Kolodiazhnyi}\ \emph {et~al.}(2010)\citenamefont
  {Kolodiazhnyi}, \citenamefont {Tachibana}, \citenamefont {Kawaji},
  \citenamefont {Hwang},\ and\ \citenamefont
  {Takayama-Muromachi}}]{kolodiazhnyi2010}%
  \BibitemOpen
  \bibfield  {author} {\bibinfo {author} {\bibfnamefont {T.}~\bibnamefont
  {Kolodiazhnyi}}, \bibinfo {author} {\bibfnamefont {M.}~\bibnamefont
  {Tachibana}}, \bibinfo {author} {\bibfnamefont {H.}~\bibnamefont {Kawaji}},
  \bibinfo {author} {\bibfnamefont {J.}~\bibnamefont {Hwang}}, \ and\ \bibinfo
  {author} {\bibfnamefont {E.}~\bibnamefont {Takayama-Muromachi}},\ }\bibfield
  {title} {\enquote {\bibinfo {title} {{Persistence of Ferroelectricity in
  ${\mathrm{BaTiO}}_{3}$ through the Insulator-Metal Transition}},}\ }\href
  {\doibase 10.1103/PhysRevLett.104.147602} {\bibfield  {journal} {\bibinfo
  {journal} {Phys. Rev. Lett.}\ }\textbf {\bibinfo {volume} {104}},\ \bibinfo
  {pages} {147602} (\bibinfo {year} {2010})}\BibitemShut {NoStop}%
\bibitem [{\citenamefont {Wang}\ \emph {et~al.}(2012)\citenamefont {Wang},
  \citenamefont {Liu}, \citenamefont {Burton}, \citenamefont {Jaswal},\ and\
  \citenamefont {Tsymbal}}]{tsymbal.2012}%
  \BibitemOpen
  \bibfield  {author} {\bibinfo {author} {\bibfnamefont {Yong}\ \bibnamefont
  {Wang}}, \bibinfo {author} {\bibfnamefont {Xiaohui}\ \bibnamefont {Liu}},
  \bibinfo {author} {\bibfnamefont {J.~D.}\ \bibnamefont {Burton}}, \bibinfo
  {author} {\bibfnamefont {Sitaram~S.}\ \bibnamefont {Jaswal}}, \ and\ \bibinfo
  {author} {\bibfnamefont {Evgeny~Y.}\ \bibnamefont {Tsymbal}},\ }\bibfield
  {title} {\enquote {\bibinfo {title} {Ferroelectric instability under screened
  coulomb interactions},}\ }\href {\doibase 10.1103/PhysRevLett.109.247601}
  {\bibfield  {journal} {\bibinfo  {journal} {Phys. Rev. Lett.}\ }\textbf
  {\bibinfo {volume} {109}},\ \bibinfo {pages} {247601} (\bibinfo {year}
  {2012})}\BibitemShut {NoStop}%
\bibitem [{\citenamefont {Wang}\ \emph {et~al.}(2019)\citenamefont {Wang},
  \citenamefont {Yang}, \citenamefont {Rischau}, \citenamefont {Xu},
  \citenamefont {Ren}, \citenamefont {Lorenz}, \citenamefont {Hemberger},
  \citenamefont {Lin},\ and\ \citenamefont {Behnia}}]{Wang2019}%
  \BibitemOpen
  \bibfield  {author} {\bibinfo {author} {\bibfnamefont {Jialu}\ \bibnamefont
  {Wang}}, \bibinfo {author} {\bibfnamefont {Liangwei}\ \bibnamefont {Yang}},
  \bibinfo {author} {\bibfnamefont {Carl~Willem}\ \bibnamefont {Rischau}},
  \bibinfo {author} {\bibfnamefont {Zhuokai}\ \bibnamefont {Xu}}, \bibinfo
  {author} {\bibfnamefont {Zhi}\ \bibnamefont {Ren}}, \bibinfo {author}
  {\bibfnamefont {Thomas}\ \bibnamefont {Lorenz}}, \bibinfo {author}
  {\bibfnamefont {Joachim}\ \bibnamefont {Hemberger}}, \bibinfo {author}
  {\bibfnamefont {Xiao}\ \bibnamefont {Lin}}, \ and\ \bibinfo {author}
  {\bibfnamefont {Kamran}\ \bibnamefont {Behnia}},\ }\bibfield  {title}
  {\enquote {\bibinfo {title} {Charge transport in a polar metal},}\ }\href
  {\doibase 10.1038/s41535-019-0200-1} {\bibfield  {journal} {\bibinfo
  {journal} {npj Quantum Materials}\ }\textbf {\bibinfo {volume} {4}},\
  \bibinfo {pages} {61} (\bibinfo {year} {2019})}\BibitemShut {NoStop}%
\bibitem [{\citenamefont {Sachdev}(1999)}]{sachdev.1999}%
  \BibitemOpen
  \bibfield  {author} {\bibinfo {author} {\bibfnamefont {S.}~\bibnamefont
  {Sachdev}},\ }\href@noop {} {\emph {\bibinfo {title} {Quantum Phase
  Transitions}}}\ (\bibinfo  {publisher} {Cambridge Universtity Press},\
  \bibinfo {year} {1999})\BibitemShut {NoStop}%
\bibitem [{\citenamefont {Coleman}\ \emph {et~al.}(2001)\citenamefont
  {Coleman}, \citenamefont {P{\'{e}}pin}, \citenamefont {Si},\ and\
  \citenamefont {Ramazashvili}}]{coleman2001}%
  \BibitemOpen
  \bibfield  {author} {\bibinfo {author} {\bibfnamefont {P}~\bibnamefont
  {Coleman}}, \bibinfo {author} {\bibfnamefont {C}~\bibnamefont {P{\'{e}}pin}},
  \bibinfo {author} {\bibfnamefont {Qimiao}\ \bibnamefont {Si}}, \ and\
  \bibinfo {author} {\bibfnamefont {R}~\bibnamefont {Ramazashvili}},\
  }\bibfield  {title} {\enquote {\bibinfo {title} {How do fermi liquids get
  heavy and die?}}\ }\href {\doibase 10.1088/0953-8984/13/35/202} {\bibfield
  {journal} {\bibinfo  {journal} {Journal of Physics: Condensed Matter}\
  }\textbf {\bibinfo {volume} {13}},\ \bibinfo {pages} {R723--R738} (\bibinfo
  {year} {2001})}\BibitemShut {NoStop}%
\bibitem [{\citenamefont {Roussev}\ and\ \citenamefont
  {Millis}(2003)}]{roussev.2003}%
  \BibitemOpen
  \bibfield  {author} {\bibinfo {author} {\bibfnamefont {R.}~\bibnamefont
  {Roussev}}\ and\ \bibinfo {author} {\bibfnamefont {A.~J.}\ \bibnamefont
  {Millis}},\ }\bibfield  {title} {\enquote {\bibinfo {title} {Theory of the
  quantum paraelectric-ferroelectric transition},}\ }\href {\doibase
  10.1103/PhysRevB.67.014105} {\bibfield  {journal} {\bibinfo  {journal} {Phys.
  Rev. B}\ }\textbf {\bibinfo {volume} {67}},\ \bibinfo {pages} {014105}
  (\bibinfo {year} {2003})}\BibitemShut {NoStop}%
\bibitem [{\citenamefont {Gor'kov}\ and\ \citenamefont
  {Melik-Barkhudarov}(1961)}]{gorkov1961}%
  \BibitemOpen
  \bibfield  {author} {\bibinfo {author} {\bibfnamefont {L.~P.}\ \bibnamefont
  {Gor'kov}}\ and\ \bibinfo {author} {\bibfnamefont {T.~K.}\ \bibnamefont
  {Melik-Barkhudarov}},\ }\bibfield  {title} {\enquote {\bibinfo {title}
  {Contribution to the theory of superconductivity in an imperfect fermi
  gas},}\ }\href@noop {} {\bibfield  {journal} {\bibinfo  {journal} {Soviet
  Physics JETP}\ }\textbf {\bibinfo {volume} {13}} (\bibinfo {year}
  {1961})}\BibitemShut {NoStop}%
\bibitem [{\citenamefont {Gor'kov}(2016)}]{gorkov2016prb}%
  \BibitemOpen
  \bibfield  {author} {\bibinfo {author} {\bibfnamefont {Lev~P.}\ \bibnamefont
  {Gor'kov}},\ }\bibfield  {title} {\enquote {\bibinfo {title} {Superconducting
  transition temperature: Interacting fermi gas and phonon mechanisms in the
  nonadiabatic regime},}\ }\href {\doibase 10.1103/PhysRevB.93.054517}
  {\bibfield  {journal} {\bibinfo  {journal} {Phys. Rev. B}\ }\textbf {\bibinfo
  {volume} {93}},\ \bibinfo {pages} {054517} (\bibinfo {year}
  {2016})}\BibitemShut {NoStop}%
\bibitem [{\citenamefont {Chubukov}\ \emph {et~al.}(2016)\citenamefont
  {Chubukov}, \citenamefont {Eremin},\ and\ \citenamefont
  {Efremov}}]{chubukov2016}%
  \BibitemOpen
  \bibfield  {author} {\bibinfo {author} {\bibfnamefont {Andrey~V.}\
  \bibnamefont {Chubukov}}, \bibinfo {author} {\bibfnamefont {Ilya}\
  \bibnamefont {Eremin}}, \ and\ \bibinfo {author} {\bibfnamefont {Dmitri~V.}\
  \bibnamefont {Efremov}},\ }\bibfield  {title} {\enquote {\bibinfo {title}
  {{Superconductivity versus bound-state formation in a two-band superconductor
  with small Fermi energy: Applications to Fe pnictides/chalcogenides and doped
  ${\mathrm{SrTiO}}_{3}$}},}\ }\href {\doibase 10.1103/PhysRevB.93.174516}
  {\bibfield  {journal} {\bibinfo  {journal} {Phys. Rev. B}\ }\textbf {\bibinfo
  {volume} {93}},\ \bibinfo {pages} {174516} (\bibinfo {year}
  {2016})}\BibitemShut {NoStop}%
\bibitem [{\citenamefont {Kleinert}\ and\ \citenamefont
  {Schakel}(2003)}]{kleinert2003}%
  \BibitemOpen
  \bibfield  {author} {\bibinfo {author} {\bibfnamefont {H.}~\bibnamefont
  {Kleinert}}\ and\ \bibinfo {author} {\bibfnamefont {Adriaan M.~J.}\
  \bibnamefont {Schakel}},\ }\bibfield  {title} {\enquote {\bibinfo {title}
  {{Gauge-Invariant Critical Exponents for the Ginzburg-Landau Model}},}\
  }\href {\doibase 10.1103/PhysRevLett.90.097001} {\bibfield  {journal}
  {\bibinfo  {journal} {Phys. Rev. Lett.}\ }\textbf {\bibinfo {volume} {90}},\
  \bibinfo {pages} {097001} (\bibinfo {year} {2003})}\BibitemShut {NoStop}%
\bibitem [{\citenamefont {Gastiasoro}\ \emph
  {et~al.}(2020{\natexlab{b}})\citenamefont {Gastiasoro}, \citenamefont
  {Ruhman},\ and\ \citenamefont {Fernandes}}]{gastiasoro2020rev}%
  \BibitemOpen
  \bibfield  {author} {\bibinfo {author} {\bibfnamefont {Maria~N.}\
  \bibnamefont {Gastiasoro}}, \bibinfo {author} {\bibfnamefont {Jonathan}\
  \bibnamefont {Ruhman}}, \ and\ \bibinfo {author} {\bibfnamefont {Rafael~M.}\
  \bibnamefont {Fernandes}},\ }\bibfield  {title} {\enquote {\bibinfo {title}
  {{Superconductivity in dilute SrTiO$_3$: A review}},}\ }\href {\doibase
  https://doi.org/10.1016/j.aop.2020.168107} {\bibfield  {journal} {\bibinfo
  {journal} {Annals of Physics}\ }\textbf {\bibinfo {volume} {417}},\ \bibinfo
  {pages} {168107} (\bibinfo {year} {2020}{\natexlab{b}})},\ \bibinfo {note}
  {eliashberg theory at 60: Strong-coupling superconductivity and
  beyond}\BibitemShut {NoStop}%
\bibitem [{\citenamefont {Ahadi}\ \emph {et~al.}(2019)\citenamefont {Ahadi},
  \citenamefont {Galletti}, \citenamefont {Li}, \citenamefont {Salmani-Rezaie},
  \citenamefont {Wu},\ and\ \citenamefont {Stemmer}}]{ahadi2019}%
  \BibitemOpen
  \bibfield  {author} {\bibinfo {author} {\bibfnamefont {Kaveh}\ \bibnamefont
  {Ahadi}}, \bibinfo {author} {\bibfnamefont {Luca}\ \bibnamefont {Galletti}},
  \bibinfo {author} {\bibfnamefont {Yuntian}\ \bibnamefont {Li}}, \bibinfo
  {author} {\bibfnamefont {Salva}\ \bibnamefont {Salmani-Rezaie}}, \bibinfo
  {author} {\bibfnamefont {Wangzhou}\ \bibnamefont {Wu}}, \ and\ \bibinfo
  {author} {\bibfnamefont {Susanne}\ \bibnamefont {Stemmer}},\ }\bibfield
  {title} {\enquote {\bibinfo {title} {{Enhancing superconductivity in
  SrTiO$_3$ films with strain}},}\ }\href {\doibase 10.1126/sciadv.aaw0120}
  {\bibfield  {journal} {\bibinfo  {journal} {Science Advances}\ }\textbf
  {\bibinfo {volume} {5}} (\bibinfo {year} {2019}),\ 10.1126/sciadv.aaw0120},\
  \Eprint
  {http://arxiv.org/abs/https://advances.sciencemag.org/content/5/4/eaaw0120.full.pdf}
  {https://advances.sciencemag.org/content/5/4/eaaw0120.full.pdf} \BibitemShut
  {NoStop}%
\bibitem [{\citenamefont {Russell}\ \emph {et~al.}(2019)\citenamefont
  {Russell}, \citenamefont {Ratcliff}, \citenamefont {Ahadi}, \citenamefont
  {Dong}, \citenamefont {Stemmer},\ and\ \citenamefont {Harter}}]{stemmer2019}%
  \BibitemOpen
  \bibfield  {author} {\bibinfo {author} {\bibfnamefont {Ryan}\ \bibnamefont
  {Russell}}, \bibinfo {author} {\bibfnamefont {Noah}\ \bibnamefont
  {Ratcliff}}, \bibinfo {author} {\bibfnamefont {Kaveh}\ \bibnamefont {Ahadi}},
  \bibinfo {author} {\bibfnamefont {Lianyang}\ \bibnamefont {Dong}}, \bibinfo
  {author} {\bibfnamefont {Susanne}\ \bibnamefont {Stemmer}}, \ and\ \bibinfo
  {author} {\bibfnamefont {John~W.}\ \bibnamefont {Harter}},\ }\bibfield
  {title} {\enquote {\bibinfo {title} {{Ferroelectric enhancement of
  superconductivity in compressively strained ${\mathrm{SrTiO}}_{3}$ films}},}\
  }\href {\doibase 10.1103/PhysRevMaterials.3.091401} {\bibfield  {journal}
  {\bibinfo  {journal} {Phys. Rev. Materials}\ }\textbf {\bibinfo {volume}
  {3}},\ \bibinfo {pages} {091401} (\bibinfo {year} {2019})}\BibitemShut
  {NoStop}%
\bibitem [{\citenamefont {Herrera}\ \emph {et~al.}(2019)\citenamefont
  {Herrera}, \citenamefont {Cerbin}, \citenamefont {Jayakody}, \citenamefont
  {Dunnett}, \citenamefont {Balatsky},\ and\ \citenamefont
  {Sochnikov}}]{sochnikov2019}%
  \BibitemOpen
  \bibfield  {author} {\bibinfo {author} {\bibfnamefont {Chloe}\ \bibnamefont
  {Herrera}}, \bibinfo {author} {\bibfnamefont {Jonah}\ \bibnamefont {Cerbin}},
  \bibinfo {author} {\bibfnamefont {Amani}\ \bibnamefont {Jayakody}}, \bibinfo
  {author} {\bibfnamefont {Kirsty}\ \bibnamefont {Dunnett}}, \bibinfo {author}
  {\bibfnamefont {Alexander~V.}\ \bibnamefont {Balatsky}}, \ and\ \bibinfo
  {author} {\bibfnamefont {Ilya}\ \bibnamefont {Sochnikov}},\ }\bibfield
  {title} {\enquote {\bibinfo {title} {Strain-engineered interaction of quantum
  polar and superconducting phases},}\ }\href {\doibase
  10.1103/PhysRevMaterials.3.124801} {\bibfield  {journal} {\bibinfo  {journal}
  {Phys. Rev. Materials}\ }\textbf {\bibinfo {volume} {3}},\ \bibinfo {pages}
  {124801} (\bibinfo {year} {2019})}\BibitemShut {NoStop}%
\bibitem [{\citenamefont {Koonce}\ \emph {et~al.}(1967)\citenamefont {Koonce},
  \citenamefont {Cohen}, \citenamefont {Schooley}, \citenamefont {Hosler},\
  and\ \citenamefont {Pfeiffer}}]{koonce1967}%
  \BibitemOpen
  \bibfield  {author} {\bibinfo {author} {\bibfnamefont {C.~S.}\ \bibnamefont
  {Koonce}}, \bibinfo {author} {\bibfnamefont {Marvin~L.}\ \bibnamefont
  {Cohen}}, \bibinfo {author} {\bibfnamefont {J.~F.}\ \bibnamefont {Schooley}},
  \bibinfo {author} {\bibfnamefont {W.~R.}\ \bibnamefont {Hosler}}, \ and\
  \bibinfo {author} {\bibfnamefont {E.~R.}\ \bibnamefont {Pfeiffer}},\
  }\bibfield  {title} {\enquote {\bibinfo {title} {{Superconducting Transition
  Temperatures of Semiconducting SrTi${\mathrm{O}}_{3}$}},}\ }\href {\doibase
  10.1103/PhysRev.163.380} {\bibfield  {journal} {\bibinfo  {journal} {Phys.
  Rev.}\ }\textbf {\bibinfo {volume} {163}},\ \bibinfo {pages} {380--390}
  (\bibinfo {year} {1967})}\BibitemShut {NoStop}%
\bibitem [{\citenamefont {Collignon}\ \emph {et~al.}(2017)\citenamefont
  {Collignon}, \citenamefont {Fauqu\'e}, \citenamefont {Cavanna}, \citenamefont
  {Gennser}, \citenamefont {Mailly},\ and\ \citenamefont
  {Behnia}}]{collignon2017}%
  \BibitemOpen
  \bibfield  {author} {\bibinfo {author} {\bibfnamefont {Cl\'ement}\
  \bibnamefont {Collignon}}, \bibinfo {author} {\bibfnamefont {Beno\^{\i}t}\
  \bibnamefont {Fauqu\'e}}, \bibinfo {author} {\bibfnamefont {Antonella}\
  \bibnamefont {Cavanna}}, \bibinfo {author} {\bibfnamefont {Ulf}\ \bibnamefont
  {Gennser}}, \bibinfo {author} {\bibfnamefont {Dominique}\ \bibnamefont
  {Mailly}}, \ and\ \bibinfo {author} {\bibfnamefont {Kamran}\ \bibnamefont
  {Behnia}},\ }\bibfield  {title} {\enquote {\bibinfo {title} {Superfluid
  density and carrier concentration across a superconducting dome: The case of
  strontium titanate},}\ }\href {\doibase 10.1103/PhysRevB.96.224506}
  {\bibfield  {journal} {\bibinfo  {journal} {Phys. Rev. B}\ }\textbf {\bibinfo
  {volume} {96}},\ \bibinfo {pages} {224506} (\bibinfo {year}
  {2017})}\BibitemShut {NoStop}%
\bibitem [{\citenamefont {W\"olfle}\ and\ \citenamefont
  {Abrahams}(2011)}]{abrahams2011}%
  \BibitemOpen
  \bibfield  {author} {\bibinfo {author} {\bibfnamefont {Peter}\ \bibnamefont
  {W\"olfle}}\ and\ \bibinfo {author} {\bibfnamefont {Elihu}\ \bibnamefont
  {Abrahams}},\ }\bibfield  {title} {\enquote {\bibinfo {title} {Quasiparticles
  beyond the fermi liquid and heavy fermion criticality},}\ }\href {\doibase
  10.1103/PhysRevB.84.041101} {\bibfield  {journal} {\bibinfo  {journal} {Phys.
  Rev. B}\ }\textbf {\bibinfo {volume} {84}},\ \bibinfo {pages} {041101}
  (\bibinfo {year} {2011})}\BibitemShut {NoStop}%
\bibitem [{\citenamefont {Abrahams}\ \emph {et~al.}(2014)\citenamefont
  {Abrahams}, \citenamefont {Schmalian},\ and\ \citenamefont
  {W\"olfle}}]{abrahams2014}%
  \BibitemOpen
  \bibfield  {author} {\bibinfo {author} {\bibfnamefont {Elihu}\ \bibnamefont
  {Abrahams}}, \bibinfo {author} {\bibfnamefont {J\"org}\ \bibnamefont
  {Schmalian}}, \ and\ \bibinfo {author} {\bibfnamefont {Peter}\ \bibnamefont
  {W\"olfle}},\ }\bibfield  {title} {\enquote {\bibinfo {title}
  {Strong-coupling theory of heavy-fermion criticality},}\ }\href {\doibase
  10.1103/PhysRevB.90.045105} {\bibfield  {journal} {\bibinfo  {journal} {Phys.
  Rev. B}\ }\textbf {\bibinfo {volume} {90}},\ \bibinfo {pages} {045105}
  (\bibinfo {year} {2014})}\BibitemShut {NoStop}%
\end{thebibliography}%

\end{document}